\documentclass[11pt,twoside]{article}  
\usepackage{asp2006}
\usepackage{adassconf}

\begin{document}   

%
%

\paperID{7A.1}

%

\title{Optimal DN encoding for CCD detectors}

%
%
%
%
%

\markboth{Seaman, White \& Pence}{Optimal DN encoding for CCD detectors}

%
%
%
%

\author{R.\ L.\ Seaman}
\affil{National Optical Astronomy Observatory, Tucson, AZ, USA}

\author{R.\ L.\ White}
\affil{Space Telescope Science Institute, Baltimore, MD, USA} 

\author{W.\ D.\ Pence}
\affil{NASA Goddard Space Flight Center, Greenbelt, MD, USA}

%

\contact{Rob Seaman}
\email{seaman@noao.edu}

%
%
%

\paindex{Seaman, R.}

%



\begin{abstract}          
Image compression has been a frequent topic of presentations at ADASS.
Compression is often viewed as just a technique to fit more data into
a smaller space.  Rather, the packing of data -- its ``density'' -- affects
every facet of local data handling, long distance data transport, and
the end-to-end throughput of workflows.  In short, compression is one
aspect of proper data structuring.  For example, with FITS tile
compression the efficient representation of data is combined with an
expressive logistical paradigm for its manipulation.
\par
A deeper question remains.  Not just how best to represent the data,
but which data to represent.  CCDs are linear devices.  What does this
mean?  One thing it does not mean is that the analog-to-digital
conversion of pixels must be stored using linear data numbers (DN).
An alternative strategy of using non-linear
representations is presented, with one motivation being to magnify the
efficiency of numerical compression algorithms such as Rice.
\end{abstract}

%
%

\section{Data Representation and Compression}
The Rice compression algorithm (Rice {\em et\ al.} 1993) is
particularly familiar to astronomers in combination with the FITS Tiled
Image Convention (White {\em et\ al.} 2006, Seaman {\em et\ al.} 2007).
It has been informally paired in the past with non-linear data representations
(Nieto-Santisteban {\em et\ al.} 1999, Nicula {\em et\ al.} 2005).
Mention of compression is often
followed immediately by the word ``scheme'', as if this branch of computer
science were suitable only for black-box heuristics arrived at through a
process of trial-and-error.  This paper seeks to begin a process of layering
the issue of optimal data encoding onto a more formal foundation.

As discussed in Pence, Seaman \& White (2009A,B), the compression achieved
for an astronomical image is typically determined almost entirely by its
background noise.  Any processing that quiets the background will increase
the compression ratio.  This is the foundation for all lossy compression
algorithms -- in effect, lossy compression combines a preprocessing step
that tempers an image's noise characteristics with the subsequent
application of a lossless algorithm.

Janesick (2001) describes a non-linear hardware or software
component called a ``square-rooter''.  Instead of using a linear
analog-to-digital conversion when reading out a CCD, the square-root of
these values acts to linearize the Poisson statistics
that characterizes these detectors.  Such a DN encoding has been used on
several space missions to reduce bandwidth requirements for data transport.
While technically lossy if the square-root transform
occurs after the A/D conversion, it is equally reasonable to consider this
a type of lossless encoding since the transform
maintains oversampling of the noise at both the high and low end
of the dynamic range.

Square-root encoding itself acts as a form of compression -- 65,636
linear data levels (for 16-bit pixels) will turn into a scant 256 levels
after the square-root.  This is not nearly as dramatic when expressed in
bits, since mapping 16 bits into 8 bits corresponds to a compression ratio
of just $R = 2$.  The more important aspect, rather, is to linearize the
noise.

\section{Variance Stabilization}
A vector of random variables ({\it ie.}, an image) is said to be
{\em heteroscedastic} if the variance depends on the signal.  Many
astronomical detectors are governed by photon shot-noise, described
by Poisson statistics with the noise varying as the square-root of
the signal.  Bright pixels are noisy pixels.
 
Many familiar statistical techniques such as least-squares
fitting formally require {\em homoscedasticity}.  The penalty for
ignoring this requirement for a particular purpose may range from
negligible to significant, but the broader point is that
astronomical noise models are often very non-linear.  Data
compression techniques as applied to astronomical data must take this
into account.

Techniques exist to stabilize the variance, that is, to convert a
heteroscedastic data set to one that is homoscedastic.  One such technique
is the Anscombe transform (Anscombe 1948):
\begin{equation}
S_{gaussian} = 2 \sqrt{S_{poisson} + 3/8}
\end{equation}
The Anscombe transform will convert a Poisson sample to have (nearly)
Gaussian statistics.  The factor of 2 ensures a unit variance.

Real astronomical data does not follow a pure Poisson noise model.  For
instance, each pixel on a CCD has additive Gaussian read-noise as well
as Poisson shot-noise.  The Generalized Anscombe Transform
(Murtagh {\em et\ al.} 1995) will stabilize the variance in this case:
\begin{equation}
I'_{xy} = \frac{2}{\alpha} \sqrt {\alpha I_{xy} + \frac{3}{8} \alpha^{2} + \sigma^{2} - \alpha \overline{I}}
\end{equation}
where $\alpha = 1/gain$.  This can be rearranged into a form that uses
familiar CCD terminology
(with care to keep units straight between DN and $e^{-}$):
\begin{equation}
I'_{xy} = 2 \sqrt {gain \times (I_{xy} - bias) + \sigma_{read}^{2} + 3/8}
\end{equation}
Here the quantities have their usual meaning:
\begin{itemize}
\item $I_{xy}$ is a pixel value in analog-to-digital data numbers (DN)
\item the CCD $bias$ is in DN
\item the CCD $gain$ is in $e^{-}/DN$
\item the readout noise, $\sigma_{read}$, is in electrons, $e^{-}$
\end{itemize}

\section{Photon Transfer}
In this form, variance stabilization becomes very reminiscent of the
CCD photon transfer technique (Janesick 2007).  (Many of this paper's
conclusions will be applicable to other types of astronomical detectors,
particularly in the optical and infrared.)  Gain is the slope of a
CCD's mean-variance relation, but photon transfer permits much finer
grained analysis than simply assuming a scalar gain.

Bright pixels greatly oversample the noise.  All digital devices must
contend with quantization noise as a term like $\sqrt{1/12} = 0.2887$ DN
(Widrow \& Koll\'{a}r 2008).
To avoid stairstep aliasing effects, the
gain of a CCD should be kept lower than the read-noise, typically a value
of a few.  On the other hand, the high end of the dynamic range is
governed by shot noise, scaling as the square-root of the signal
(in $e^{-}$).  For a gain fixed at unity, the noise for pixels near
the top of the 16-bit range will be 256 DN, fully 2 orders of
magnitude above the read-noise.

Noise adds in quadrature -- permitting the gain to be scaled with total noise:
\begin{equation}
gain \propto \sqrt{\sigma_{read}^{2} + \sigma_{shot}^{2} + \sigma_{FPN}^{2}}
\end{equation}
This is simply another way to look at variance stabilization (see, for
example, figure 11.19 from Janesick 2007).  The goal is to tame all the
sources of noise.

Astronomical images are not simply random agglomerations of pixels, they
result from astronomical detectors with specific characteristics and
noise models.  For CCDs (quantum yield $\eta = 1$) this becomes:
\begin{equation}
gain \propto \sqrt{\sigma_{read}^{2} + \eta S + (P_{N} S)^{2}}
\end{equation}
This presents a quandry, since the fixed pattern noise (FPN) dominates
both the Gaussian read-noise and Poisson shot noise at the bright end
(if not flatfielded).

\section{Thoughts on Optimal Encoding}
The question is how to put this together into a coherent recipe.  Work
is ongoing, but broad strokes are clear.  First, the quandry is only
apparent, since to first order the signal (and thus the FPN) is negligible.
Recall that
the compression ratio, {\em R}, is determined
by the noise in the {\em background} (Pence {\em et\ al.} 2009A,B):
\begin{equation}
R = \frac{\mbox{\small\em BITPIX}}{N_{bits} + K}
\end{equation}
where ($\sigma$ estimated per Stoehr 2007):
\begin{equation}
N_{bits} = \log_{2}\sigma + 1.792
\end{equation}

Second, phenomenological schemes such as eq. 7.86 from Janesick (2001):
\begin{equation}
DN_{out} = \sqrt{DN_{in}} \times 2^{\:(N_{out} - \mbox{\scriptsize\em BITPIX}/2)}
\end{equation}
are simplified versions of the more formal functional behavior.
Equation (8) is the same as the Anscombe transform
when $N_{out} = 1 + \mbox{\scriptsize\em BITPIX}/2$
({\em eg.}, $N_{out} = 9$ for {\scriptsize\em BITPIX} = 16)
and when the 3/8 term is negligible ($DN_{in} > \sim 30$).

Whatever functional form, a lookup table is an efficient implementation.
The mapping $DN_{in} \Rightarrow DN_{out}$ is surjective; several inputs
map to one output.  The inverse LUT is compact as a header or table
structure under the FITS Tiled Image Convention.  Libraries like CFITSIO
(http://heasarc.gsfc.nasa.gov/fitsio) that support tiling will be able to
use such LUTs to transparently recover the linear data with adequate noise
sampling.
For some purposes linearization is not needed.  Square root encoding
may be just what is wanted to load an image display, or for homoscedastic
statistics ({\em eg.}, principle components analysis), or for multiresolution
techniques using wavelets (Murtagh {\em et\ al.} 1995).

Some issues remain to be resolved:
\begin{enumerate}
\item Since it is the background noise that matters, the low end mapping
is key.  However, in equation (3) each pixel value is corrected for the
bias, potentially driving the radicand negative.
How can this best be resolved?
\item What is the best high end mapping for raw and flat-fielded images?
\item When bringing the CCD noise model into alignment with the
Generalized Anscombe Transform, it is easy to get lost between
$e^{-}$ and DNs (also, what most call the {\em gain}, Janesick calls
{\em sensitivity}, the inverse gain).  Effects like quantum yield
need to be folded into the variance stabilization.
\item The Data Compression literature ({\em eg.}, Salomon 2004) is built
on a foundation of Information Theory ({\em eg.}, Cover \& Thomas 2006).
As such, it may be revealing to recast both the variance stabilization
and CCD noise modeling techniques in terms of the Shannon entropy (1948).
\end{enumerate}


\begin{references}

\reference Anscombe, F.\ J.\ 1948, Biometrika, 15, 246
\reference Cover, T.\ M. \& Thomas, J.\ A.\ 2006,``Elements of Information Theory'' (Wiley-Interscience)
\reference Janesick, J.\ 2001, ``Scientific Charge Coupled Devices'' (SPIE Press)
\reference Janesick, J.\ 2007, ``Photon Transfer, DN $\rightarrow \lambda$'' (SPIE Press)
\reference Murtagh, F., Starck, J.-L.\ \& Bijaoui, A.\ 1995, \aaps, 112, 179
\reference Nicula, B., Berghmans, D.\ \& Hochedez, J-F.\ 2005, Solar Physics, 228, 253
\reference Nieto-Santisteban, M.\ et\ al.\ 1999, \adassviii, 137
\reference Pence, W., Seaman, R.\ \& White, R.\ 2009A, ``Lossless Astronomical Image Compression and the Effects of Noise'', \pasp, in press
\reference Pence, W., Seaman, R.\ \& White, R.\ 2009B, \adassxviii, \paperref{O2.1}
\reference Rice, R.\ F., Yeh, P.-S.\ \& Miller, W.\ H.\ 1993, in Proc. 9th AIAA Comp. in Aerospace Conf., (AIAA-93-4541-CP), Amer. Inst. of Aeronautics \& Astronautics
\reference Salomon, D.\ 2004, ``Data Compression: The Complete Reference, 3rd Ed.'' (Springer)
\reference Seaman, R., et\ al.\ 2007, \adassxvi, 483
\reference Shannon, C.\ E.\ 1948, ``A Mathematical Theory of Communication'', Bell System Technical Journal, 27, 379 \& 623
\reference Stoehr, F. June 2007, ST-ECF Newsletter
\reference White, R. et\ al.\ 2006, ``Tiled Image Convention for Storing Compressed Images in FITS Binary Tables'', http://fits.gsfc.nasa.gov/registry/tilecompression/tilecompression.pdf
\reference Widrow, B.\ \& Koll\'{a}r, I.\ 2008, ``Quantization Noise'' (Cambridge University Press)

\end{references}
\end{document}